\documentclass[doublespacing]{elsart}

\usepackage{amssymb}
\usepackage{epsfig}
\usepackage{graphicx}

\begin{document}

\begin{frontmatter}



\title{Preferential attachment with information filtering - \\
node degree probability distribution properties}% Force line breaks with \\


\author[irb]{Hrvoje \v Stefan\v ci\' c}
\ead{shrvoje@thphys.irb.hr}
\author[irb]{Vinko Zlati\' c}%
\ead{vzlatic@irb.hr}

\address[irb]{Theoretical Physics Division, Rudjer Bo\v skovi\' c Institute  \\
P.O.B. 180, HR-10002 Zagreb, Croatia }

\begin{abstract}
A network growth mechanism based on a two-step preferential rule is investigated
as a model of network growth in which no global knowledge of the network is
required.
In the first filtering step a subset of fixed size $m$ of existing nodes is
randomly chosen. In the second
step the preferential rule of attachment is applied to the chosen subset. The
characteristics of thus formed networks are explored using two approaches: computer
simulations of network growth and a theoretical description based on a master
equation. The results of the two approaches are in excellent agreement. Special
emphasis is put on the investigation of the node degree probability distribution. It
is found that the tail of the distribution has the exponential form given by
$exp(-k/m)$. Implications of the node degree distribution with such tail
characteristics are briefly discussed.
\end{abstract}

\begin{keyword}
Scale-free networks \sep Master equation \sep Degree distribution

\PACS 89.75.Hc \sep 89.75.Da \sep 05.10.Gg \sep 02.50.Ng

%\pacs{89.75.Hc, 89.75.Da, 05.10.Gg, 02.50.Ng}% PACS, the Physics and Astronomy


\end{keyword}
\end{frontmatter}

\section{Introduction}

In many aspects of the society and the world we live in, network
structures are a common occurrence. At the level of their
phenomenological description, these networks are very
heterogeneous both in the nature of their nodes and their links
and, apart from their very network nature, seemingly have nothing
in common. However, more detailed statistical analysis of various
communication, transport, social, biological and other networks
have revealed amazing similarities in their statistical
characteristics. These findings, implying the existence of a
general underlying organizing principle, have stimulated an
entirely new wave of research and have given birth to the field of
complex networks \cite{AB02,DM02,N03,WS98}.

The statistical characteristic of complex networks that proved to be the most
suitable for demonstrating the common features of very different networks is the
distribution of node degrees. It was found that in complex networks as different
as Internet \cite{Chen02}, world wide web \cite{Broder00}, citations \cite{Redner98}, scientific
collaborations \cite{GI95}, metabolic interactions \cite{Jeong01} and many others, the
distribution of degrees has a power law character for large values of
node degrees, i.e. a power law tail. This property of the
aforementioned networks, which are also referred to
as {\em scale-free} networks,  was shown to be important in dynamical processes
on networks, such as human
epidemics and e-mail virus spreading \cite{PV01a,PV01b} and resilience to Internet
attacks \cite{AJB00,Cruciti03}. A deeper understanding and modeling of laws behind the
formation of scale-free networks was clearly necessary.

The first crucial step towards unraveling the principles behind ``scale-free"
networks was undertaken by Barab\'{a}si and Albert \cite{BA99b}. Their model
(in further text referred to as BA model) is
founded on two simple, but essential principles \cite{BAJ99}: network growth and
preferential attachment. The network growth is realized by adding a new
node to the network at each time step. The new node attaches to $r$ already
existing nodes in the network. The rule of preferential attachment dictates
that the probability of attaching a new node to the existing node is
proportional to the degree of the existing node. These two basic principles
are sufficient to reproduce a power law tail. A host of modifications and
elaborations of the preferential attachment rule in the evolving networks followed
very soon after the formulation of the BA model \cite{BB01a,BB01b,KR01,KRL00}. These
models succeeded in reproducing some of the statistical characteristics of complex
networks, such as the power law exponent, in a more realistic manner.

Apart from the preferential attachment rule, it is possible to question how other
elements of the BA model could be modified to better capture the properties of real world
networks. In the process of adding a new node to a very large network (such as www),
it is reasonable to expect that only part of information about the entire network
is effectively available. Such an assumption is grounded in the fact that nodes usually have
a limited processing capability or have (e.g. topical) interest only in a part of the
network. It is also reasonable to expect that within the processed part of
the network the attachment rule of the new node to the old ones is not completely
random, but is preferential. Namely, some of the old nodes are more ``appealing" to
the new ones.
%An implicit assumption behind the
%preferential attachment rule is the access to the information (in the BA model
%case, the node degrees) of the entire network. For very large networks, such as
%communication networks like Internet or www, this assumption of the global
%knowledge on network structure is clearly unrealistic. Therefore, it is
%reasonable to consider the variants of BA model which relax the assumption of
%the global knowledge of the network statistical features.
A reasonable variant which encompasses the aforementioned properties
is the filtration of global information \cite{mossa}. In this model
of network growth, a new node selects a subset of existing nodes and then
applies the rule of preferential attachment to the selected subset. The
principle of selection may be chosen in various ways. The subset may be formed
by randomly choosing a finite fraction or a fixed number of existing nodes
\cite{mossa}. In the case of the choice of the finite fraction, however, the
selected subset may become unrealistically large for large networks and moderate
fractions. One can also introduce an additional attribute of every node,
and select into the subset only those existing nodes the attributes of which are
sufficiently similar to the attribute of the new node \cite{esp}. This model
may, e.g. simulate the attachment to topically similar pages on the www.

The investigation of the model of preferential attachment based on the realistic
access to the information on the network structure is also important with
respect to the properties of dynamical processes on networks. It has been shown
\cite{PV01a} that in scale-free networks there exists no threshold for the spreading
of epidemics in human society or computer viruses in the Internet. The key feature for the
nonexistence of the epidemics spreading threshold (or equivalently, lack of the
phase transition) is the power law tail of the node degree distribution in
scale-free networks. In the realistic modifications of information filtering
it is reasonable to expect that some of the characteristics of the node degree
distribution might change. One of the aims of this paper is to determine
the behavior of the node degree distribution function for large degrees and its
consequences for the dynamical processes on networks formed by information
filtering.

In this paper, we study a particular model of information filtering with
special emphasis on its theoretical description and tail
characteristics. In this model, already introduced in \cite{mossa}, a selected
subset, which we call the filtration subset,
 is formed by randomly choosing a fixed number of existing nodes. We
concentrate on the determination of the node degree distribution and approach
the problem in two different ways: theoretically, applying the master equation
approach of Dorogovtsev and Mendes \cite{DMS00} and using computer simulations, in a
manner analogous to \cite{mossa}. The results of these two approaches are
compared and their agreement discussed in detail.

\section{Model}

\begin{figure}[h]
\centerline{%\rotatebox{-90}
{\resizebox{0.6\textwidth}{!}
{\includegraphics{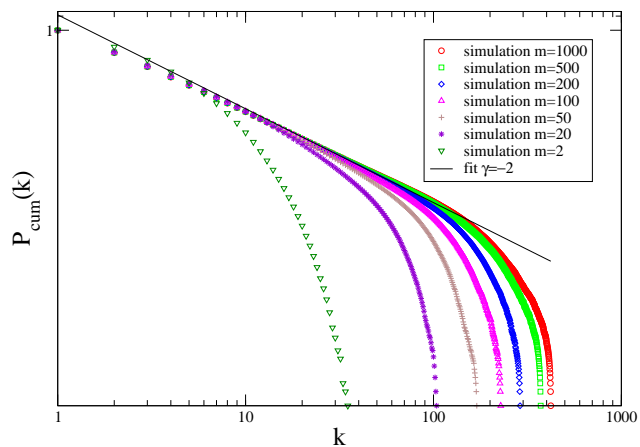}}}}
\caption{\label{fig:1} The node degree cumulative probability distribution
obtained in the simulations of the network growth is displayed for the values of
the filtering subset $m=2, 20, 50, 100, 200, 500$ and $1000$.
The cumulative probability
distributions follow the power law for a range of node degrees, whereas for large
values they exhibit a quickly decaying tail. The straight line represents the fit of
the power law behavior with the exponent -2.}
\end{figure}

We start with the discussion of computer simulations of the network growth in
our model. The network grows by addition of one new node at each time step.
Each new node connects to the existing network with {\em one} link. This special
choice facilitates the theoretical treatment and ensures a higher level of
analytical tractability. Therefore, at the time step $t$, there are $t$ links in
the network. In every simulation of the network growth, a core of the network is
initially formed. The core of the network is formed in a growth procedure where each
new node is randomly connected to one of the existing nodes. The size of the
network core is chosen to be larger than the maximal size of the
filtration subset
that is considered in this paper. As we consider filtration subset sizes up to
1000, the network core size is taken to be 1100 in all simulations. The nodes
are numbered starting from 0, and the network is allowed to grow until the
number of nodes, and respectively the number of links, reaches the maximal value
$n_{max}$, which is generally taken as $n_{max}=10^{6}$. The statistical
descriptors of the investigated network are obtained as averages over 100
independent realizations of the network topology. The stability of the node
degree probability distribution was tested by varying the final network
size from $10^{5}$ to $10^{7}$. The simulations with different values of
$n_{max}$ show the stability of the node degree probability distribution
%\ref{fig:asymptote}
 for all but the largest node degrees. The probability distributions are
characterized by the exponential cut-off in the tail. For larger $n_{max}$, the
position of the cut-off is shifted more towards larger node degrees, as already found
in \cite{mossa}. The stability issues will further be addressed in section
\ref{results}.
% which confirms the stability of the node degree probability
%distribution, as already found in \cite{mossa}.
It has also been found that the
node degree probability distribution does not depend in any significant way on
the choice of the core of the simulated network. The results of simulations are
displayed in Fig. \ref{fig:1}. The cumulative probability distribution clearly
exhibits the power law scaling for a broad range of node degrees and has a
rapidly falling tail. The exponent of the power law behavior of the cumulative
probability distribution is close to -2.

\section{Theoretical treatment}

\label{theor}

In the theoretical treatment of the node degree distributions we adopt the
master equation approach of \cite{DMS00}

In the type of the growing network
investigated in this paper, each node is uniquely identified by the time of its
attachment to the network which is hereafter denoted by $s$. In such a
framework, we introduce the probability that a node created at time $s$, at some
time $t$, where $t \ge s$, has the degree $k$ and denote it as $p(k,s,t)$.
The probability that a randomly chosen node has a degree $k$ at time $t$ is
denoted by $P(k,t)$. The cumulative probability distribution is denoted by
$P_{cum}(k,t)$ and defined as $P_{cum}(k,t)=\sum_{l=k}^\infty P(l,t)$.
The quantity of interest in our theoretical approach is the probability that
the node added at time $t$ attaches to the node created at time $s$ which at
that time has the degree $k$. The size of the filtration subset is $m$. The
probability that the chosen filtration
subset at time $t$ contains the node created at $s$
with the degree $k$ and that the sum of degrees of all nodes in the subset
equals $l$ is

\begin{equation}
\label{1}
    w(l-k,m,t)=\frac{
    \left(\begin{array}{c}
    t-1 \\ m-1
    \end{array}\right)\cdot
    g(l-k,t)}
    {
    \left(\begin{array}{c}
    t \\ m
    \end{array}\right)} \, ,
\end{equation}

where $g(u,t)$ is the probability that the sum of degrees of $m-1$ nodes is $u$
at time $t$. It is obtained as a convolution of the order $m-1$ of the
probability function $P(k,t-1)$, i.e.

\begin{equation}
\label{2} g(u,t)=P^{*(m-1)}(u,t-1) \, .
\end{equation}

The expression (\ref{2}) allows us to write equation (\ref{1}) as

\begin{equation}
\label{3} w(l-k,m,t)=\frac{m}{t} P^{*(m-1)}(l-k,t-1) \, .
\end{equation}

This probability of choosing the desired node in a selected
filtration subset must be
multiplied by the probability of choosing the node created at time $s$ from the
selected filtration
set. This phase of selection is preferential, i.e. the probability for
any node being chosen is proportional to its degree. For the selected subset
described above, this probability amounts to $k/l$. With all necessary
elements specified, we can formulate the master equation for the probabilities
$p(k,s,t)$:

\begin{eqnarray}
\label{4} p(k,s,t)& = & \left( \sum_{l=k-1+m-1}^{k-1+a_{max}}
w(l-(k-1),m,t)\frac{k-1}{l} \right) \nonumber \\
& \times & p(k-1,s,t-1) \nonumber \\
&+& \left(1-\sum_{l=k+m-1}^{k+a_{max}}
w(l-k,m,t)\frac{k}{l}\right) \nonumber \\
&\times& p(k,s,t-1) \, ,
\end{eqnarray}

for $k \ge 2$ and

\begin{eqnarray}
\label{19}
p(1,s,t)&=&\delta_{s,t}+(1-\delta_{s,t})\left(1-\sum_{l=m}^{1+a_{max}}
w(l-1,m,t)\frac{1}{l}\right) \nonumber \\
&\times& p(1,s,t-1) \, .
\end{eqnarray}

The limits of summation take into account the fact that the minimal degree of
any node in the network is 1. The term in the upper limit of the summations,
denoted by $a_{max}$, specifies the maximal value the sum of $m-1$ degrees can
have. In the limit of infinitely large networks, it becomes infinite.
Performing index substitutions in the summations on the right-hand side of
(\ref{4}) and (\ref{19}) and using (\ref{3}), we obtain

\begin{eqnarray}
\label{9} p(k,s,t)&=&\frac{k-1}{t}\left(m\sum_{r=m-1}^{a_{max}}
\frac{P^{*(m-1)}(r,t-1)}{r+k-1}\right) \nonumber \\
&\times& p(k-1,s,t-1)\nonumber \\
&+& \left(1-\frac{k}{t}\left(m\sum_{r=m-1}^{a_{max}}
\frac{P^{*(m-1)}(r,t-1)}{r+k}\right)\right) \nonumber \\
&\times&p(k,s,t-1)
\end{eqnarray}

and

\begin{eqnarray}
\label{22}
p(1,s,t)&=&\delta_{s,t}+\left(1-\frac{m}{t}\sum_{r=m-1}^{a_{max}}
\frac{P^{*(m-1)}(r,t-1)}{r+1}\right) \nonumber \\
&\times& p(1,s,t-1) \, .
\end{eqnarray}

The probability function for the node degree can be further defined in terms of
the probability $p(k,s,t)$ as

\begin{equation}
\label{10} P(k,t)=\frac{1}{t+1} \sum_{s=0}^{t}p(k,s,t) \, .
\end{equation}

Furthermore, we introduce a function

\begin{equation}
\label{13}
f_m(k,t)=m\sum_{r=m-1}^{a_{max}}\frac{P^{*(m-1)}(r,t-1)}{r+k} \, ,
\end{equation}

which, together with (\ref{10}), leads to the recursive relations for the node
degree probability function:

\begin{eqnarray}
\label{14}
P(k,t)&=&\frac{k-1}{t-1}f_m
(k-1,t)P(k-1,t-1) \nonumber \\
&+& \left(1-\frac{k}{t}f_m (k,t)\right)\frac{t}{t+1}P(k,t-1)
\end{eqnarray}

and

\begin{equation}
\label{25}
P(1,t)=\frac{1}{t+1}+\left(1-\frac{1}{t}f_m(1,t)\right)\frac{t}{t+1}P(1,t-1)
\, .
\end{equation}

The set of equations displayed so far operates with the time-dependent
probability distributions. As the results of simulations imply, the node degree
probability distribution becomes stable when the growing network becomes large
enough. Therefore, in further considerations we investigate the stable
probability distribution. This implies that $ P(k,t)\rightarrow P(k)$ and
$f_m (k,t) \rightarrow f_m(k)$. For the stable degree node probability
distribution, we have the following relations:

\begin{equation}
\label{17}
P(k)=\frac{(k-1)f_m (k-1) P(k-1)}{1+k f_m(k)}
\end{equation}

and

\begin{equation}
\label{27}
P(1) = \frac{1}{1+f_m (1)} \, .
\end{equation}

The set of equations (\ref{17}) and (\ref{27}) is a complicated set of equations
given the fact that the functions $f_m(k)$ depend on the entire probability
distribution $P(k)$ in a nonlinear fashion. In solving this set of equations we adopt
the following iterative procedure. The function $f_m(k)$ is calculated using a guess
initial node degree probability distribution of the form $P(k) \sim k^{-\gamma}$ with
$\gamma = 3$ (reminiscent of the Albert-Barab\'{a}si model). The function $f_m(k)$
thus obtained, is further used to recursively calculate the probability distribution
$P(k)$ from the set of equations (\ref{17}) and  (\ref{27}). The obtained probability
distribution is then used to calculate the function $f_{m}(k)$ and the entire
procedure is repeated until satisfactory convergence is achieved.

\section{Results}

\label{results}

\begin{figure}
\centerline{%\rotatebox{-90}
{\resizebox{0.6\textwidth}{!}
{\includegraphics{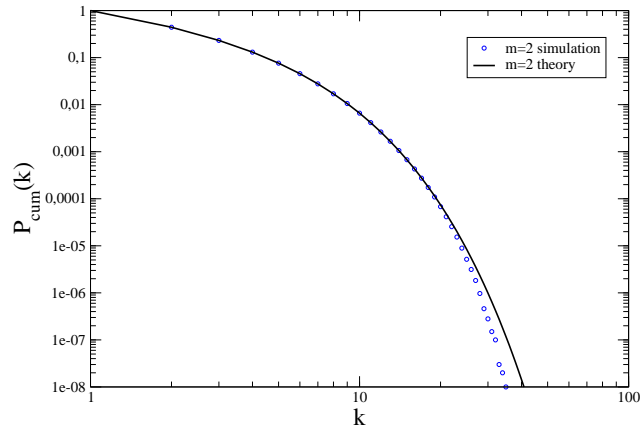}}}}
\caption{\label{fig:2a} The comparison of the node degree cumulative
probability distributions obtained by simulation (circles) and theoretical
calculations (full line) for $m=2$. Theoretical and simulational distributions are in
excellent agreement although the power law aspect of the
distribution is very weakly expressed. The disagreement in the tail
of the distributions may be attributed to the finite size effect in the
simulational data.}
\end{figure}

\begin{figure}
\centerline{%\rotatebox{-90}
{\resizebox{0.6\textwidth}{!}
{\includegraphics{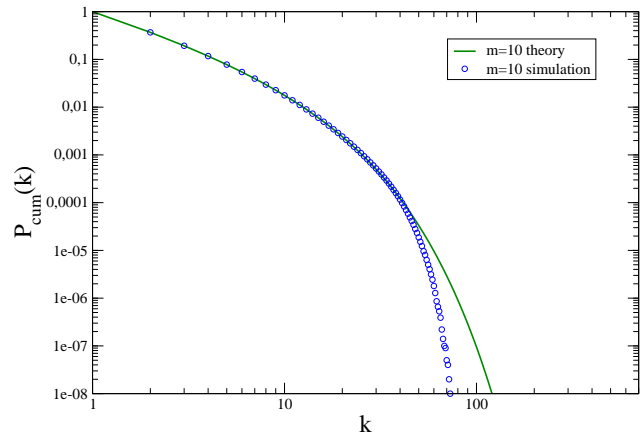}}}}
\caption{\label{fig:2b} The comparison of the node degree cumulative
probability distributions obtained by simulation (circles) and theoretical
calculations (full line) for $m=10$. Excellent agreement of the two distributions
is evident. The disagreement at the largest node degrees is attributed to the
final size effects in the simulational distribution.   }
\end{figure}

\begin{figure}
\centerline{%\rotatebox{-90}
{\resizebox{0.6\textwidth}{!}
{\includegraphics{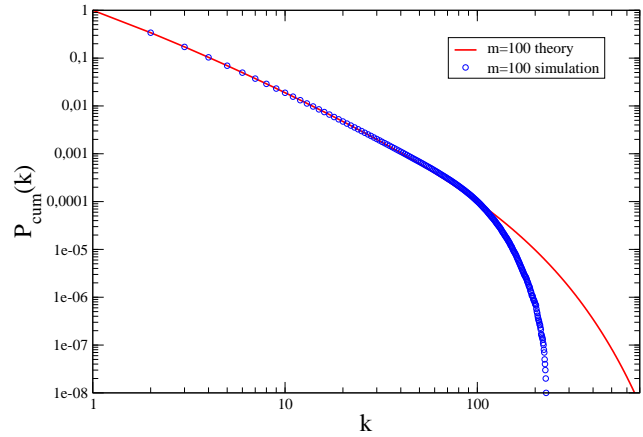}}}}
\caption{\label{fig:2c} The comparison of the node degree cumulative
probability distributions obtained by simulation (circles) and theoretical
calculations (full line) for $m=100$. The two distributions are in excellent
agreement, except for the largest values of node degrees. The disagreement in this
limit is attributed to the final size of the simulational data set. }
\end{figure}

\begin{figure}
\centerline{%\rotatebox{-90}
{\resizebox{0.6\textwidth}{!}
{\includegraphics{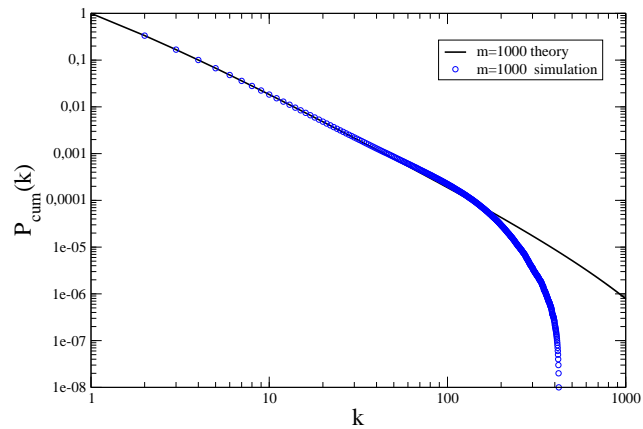}}}}
\caption{\label{fig:2d} The comparison of the node degree cumulative
probability distributions obtained by simulation (circles) and theoretical
calculations (full line) for $m=1000$. Excellent agreement of the two
distributions is absent only for the largest values of the node degrees which
is explained as a consequence of a finite data set obtained by simulation. }
\end{figure}

\begin{figure}
\centerline{%\rotatebox{-90}
{\resizebox{0.6\textwidth}{!} {\includegraphics{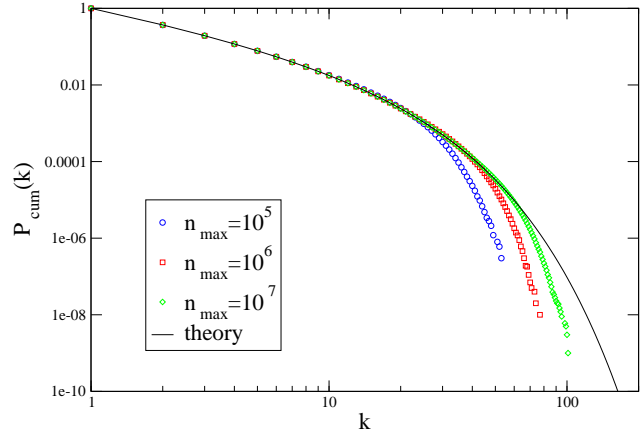}}}}
\caption{\label{fig:asymptote} The asymptotical approach of the
simulated cumulative degree probability distribution for sample
size $m=10$, and different maximum network sizes $n_{max} \in
\{10^5, 10^6, 10^7\}$ to the theoretically obtained one. Other
sample sizes $m$ also exhibit same behavior.}
\end{figure}

The results of the calculations of the theoretical node degree probability
distribution show excellent agreement with the analogous results obtained in
simulations. In Figs \ref{fig:2a}, \ref{fig:2b}, \ref{fig:2c}, and \ref{fig:2d}
we show graphs of cumulative degree node
probability distributions for the values of the subset size $m=2$, $10$, $100$, and
$1000$. The agreement of the two cumulative distributions obtained by completely
different procedures is excellent, except for the largest values of the node
degrees. The disagreement in this area of $k$ is attributed to the finite size
effects of the cumulative distribution obtained by simulation. The agreement
between the theoretical and simulational cumulative probability distribution is
present not only for the linear parts of the graphs in the log-log plots, but
also for the curved ones which represent the declination from the power law
i.e. the quickly decaying tail. This fact is especially clear for
smaller values of the filtering subset $m$ (see e.g. Figs \ref{fig:2a} and
\ref{fig:2b}) Therefore, the theoretical distribution functions are capable of
explaining the characteristics of the tail of the cumulative probability
distribution $P_{cum}(k)$. These findings support the use of the theoretically
constructed distributions as predictions of the results that would be
obtained in the simulations of larger networks. The stability of the node degree
probability distribution for $m=10$
is shown in Fig. \ref{fig:asymptote}. For larger sizes of
the simulated network $n_{max}$, the node degree probability distribution approaches
more to the theoretically obtained one. From the distributions displayed in Fig.
\ref{fig:asymptote} it is natural to expect that the simulated and theoretical curves
will be in full agreement for the infinitely large simulated network. An analogous
behavior is observed for all sample sizes $m$. The disagreements in the tails
of probability distributions between finite size simulated
networks and theory are attributed to the finite size of the simulated networks.

\begin{figure}
\centerline{%\rotatebox{-90}
{\resizebox{0.6\textwidth}{!}
{\includegraphics{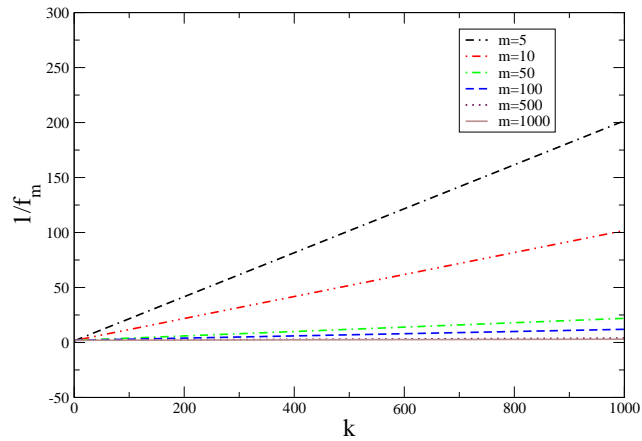}}}}
\caption{\label{fig:3} The inverse of the function $f_{m}$ as the function of
the node degree for the filtering subset sizes $m=5, 10, 50, 100, 500,$ and
$1000$. The graphs clearly show the linearity of functions $1/f_{m}$. }
\end{figure}

Inspection of the theoretically constructed cumulative probability distribution
shows that this distribution has a tail decreasing faster than the power law
which dominates the appearance of the distribution for quite a large number of
decades in $k$. Further inspection of these graphs in a linear-log plots reveals
the exponential nature of the distribution tails. Let us further examine
characteristics of these tails. Interesting information in this direction can be
provided by taking into consideration the function $f_{m}(k)$ appearing in the
recursion relation for the probability function $P(k)$. In Fig. \ref{fig:3} it
is clearly shown that $1/f_{m}$ is very linear, i.e. the form of this function
is

\begin{equation}
\label{eq:fmform}
f_{m}(k) = \frac{1}{a_{m}+b_{m}k} \, .
\end{equation}

Based on this form of dependence of $f_{m}$ on $k$, we can treat the set of
equations (\ref{17}) and (\ref{27}) in a much more analytically tractable
fashion. For large values of $k$ it is possible to write Eq. (\ref{17}) in a
continuum approximation as a differential equation

\begin{equation}
\label{29}
P(k)=-\frac{d}{dk}\left(\frac{k}{a_{m}+b_{m}k}P(k)\right) \, ,
\end{equation}

or, equivalently as

\begin{equation}
\label{31}
k\frac{dP(k)}{dk}=\left(\frac{b_{m}k}{a_{m}+b_{m}k}-a_{m}-b_{m}k-1\right)P(k) \, .
\end{equation}

The solution of Eq. (\ref{31}) is of the form

\begin{equation}
\label{33}
P(k)=C (a_{m}+b_{m}k) k^{-(a_{m}+1)}e^{-b_{m}k} \, ,
\end{equation}

where $C$ is the integration constant. The cumulative probability distribution,
obtained by the integration of expression (\ref{33}), then acquires the form

\begin{equation}
\label{33cum}
P_{cum}(k) \sim k^{-a_{m}}e^{-b_{m}k} \, .
\end{equation}

\begin{figure}
\centerline{%\rotatebox{-90}
{\resizebox{0.6\textwidth}{!}
{\includegraphics{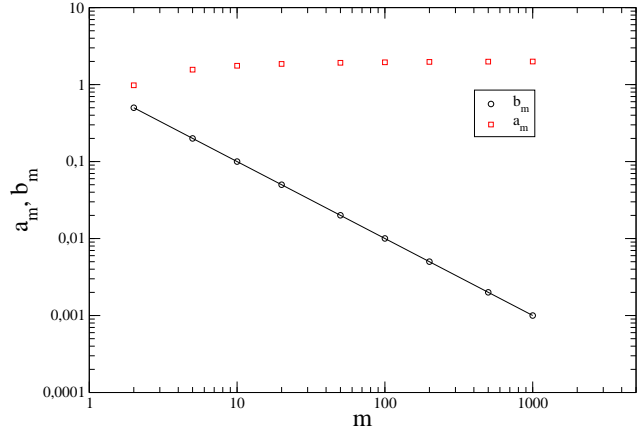}}}}
\caption{\label{fig:4} The values of the coefficients $a_{m}$ and $b_{m}$ for 9
different values of $m$. The coefficient $a_{m}$ tends to the value 2 as $m$
increases. The coefficient $b_{m}$ fits very well to the function $1/m$ as seen
from the best fit function.}
\end{figure}

\begin{figure}
\centerline{%\rotatebox{-90}
{\resizebox{0.6\textwidth}{!}
{\includegraphics{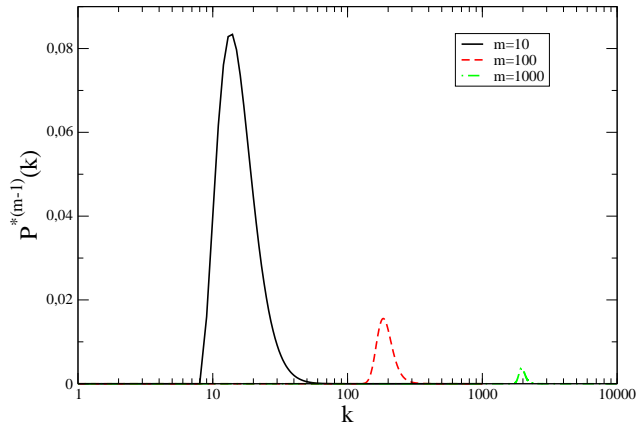}}}}
\caption{\label{fig:conv} Probability distribution $P^{*(m-1)}(k)$ displayed for
three values of $m=10, 100$, and $1000$. The localized character of the
distributions is clearly visible from the graphs.}
\end{figure}

The final result demonstrates that the tail of the distribution has the
exponential nature which is governed by the value of the coefficient $b_{m}$.
The value of the coefficient $a_{m}$ determines the exponent of the power law
part of the cumulative probability function form. Figure \ref{fig:4},
displaying the dependence of the coefficients $a_{m}$ and $b_{m}$ on $m$, shows that
for $m \ge 5$, the value of the coefficient $a_{m}$ lies in the vicinity of 2,
which produces the power law form analogous to the original Albert-Barab\'{a}si
power law. The larger the value of $m$, the closer is $a_{m}$ to the value 2.
The coefficient $b_{m}$ falls with the increase of the subset size
$m$. Its dependence on $m$ is given by the regression
$b_{m}=1.0003 \; m^{-0.99982}$. We see that a simple formula

\begin{equation}
\label{eq:bmvsm}
b_{m} = \frac{1}{m} \,
\end{equation}

describes the dependence of $b_{m}$ on $m$ very well. Let us consider this fact
in more detail to gain further understanding of the characteristics of the
function $f_{m}$ and its coefficients $a_{m}$ and $b_{m}$. The convolution
function $P^{*(m-1)}(k)$
is a rather well localized distribution function, irrespectively of the
(non)localized character of the probability distribution function $P(k)$. This
kind of behavior of the convolution probability function is illustrated in Fig.
\ref{fig:conv}. The function $f_{m}(k)$ can then be written in the form

\begin{equation}
\label{eq:fmconv}
f_m(k)=\sum_{r=m-1}^{a_{max}}\frac{P^{*(m-1)}(r)}{\frac{r}{m}+\frac{k}{m}} \, .
\end{equation}

The convolution probability distribution $P^{*(m-1)}(k)$
is localized around
its average value, amounting to $(m-1) \overline{k}$, where $\overline{k}$ is
the average node degree in the network. The largest contribution to the function
$f_{m}(k)$ comes from the interval around the average value of the convolution
probability distribution $P^{*(m-1)}(k)$. Assuming that the denominator in the
sum of the expression does not vary much in the interval where the convolution
probability distribution is non-negligible, we can approximate $r$ in the
denominator of this expression with $(m-1) \overline{k}$. Now it is possible to
trivially sum the terms in the numerator to 1 (since it is a probability
function) and the expression (\ref{eq:fmconv}) becomes

\begin{equation}
\label{eq:fmsimple}
f_m(k)=\frac{1}{\frac{(m-1) \overline{k}}{m}+\frac{k}{m}} \, .
\end{equation}

This chain of arguments validates the expression (\ref{eq:fmform})
and explains its origin. It is interesting to note, as specified
in Fig. \ref{fig:3}, that the assumptions introduced in the
preceding paragraph work satisfactorily well even for the smallest
values of the parameter $m$.

\section{Conclusion}

In conclusion, we have investigated the modification of the BA model with the fixed
size filtration subset. The node degree probability distribution was studied
using simulations of network growth and theoretical modeling. These two
approaches yield results for the node degree probability distribution which are
in excellent agreement. The theoretical approach enables one to gain a deeper insight
into the characteristics of the distribution. The cumulative probability distribution
has a power law character with an exponential tail. The power law exponent
approaches the BA model value -2 as the filtration subset size $m$ becomes
larger. Furthermore, excellent agreement of the simulational and theoretical
distributions for all node degrees except the largest, enables us to use
theoretical distribution in gaining information on the distribution tail beyond
information available from simulations which are constrained by the finite sizes
of the simulational data sets. The tail decays exponentially, the decay being
slower for the larger values of $m$. More precisely, the exponential factor is
of the form $e^{-b_{m} k}$, where $b_{m}=1/m$ is a very accurate description.
The results reported in this paper demonstrate the possible mechanism of the
realistic access to the network information in the process of growth of the
network by preferential attachment. It is clearly shown how the declination
from the concept of the global knowledge of the network leads to the
modifications in the node degree probability distribution and how the
exponential tail appears. These results have interesting implications on
the problem of epidemics spreading since the exponential tail in the node degree
probability function results in the existence of non-vanishing epidemics
threshold. All other dynamic processes on networks that are sensitive to the
(in)finiteness of the moments of the node degree probability distribution are
affected as well.

The modification of the preferential attachment rule considered in this paper is
just one of the possible elaborations towards the understanding of realistic
networks. Given the omnipresence of network
structures and their large impact on our lives and societies, further
investigations of their formation mechanism are called for.

{\bf Acknowledgments.} The authors would like to thank A. \v{S}iber for useful
comments on the manuscript. This work was supported by the Ministry of Science
and Technology of the Republic of Croatia under the contract numbers 0098002 and
0098004.

\appendix

\section{Figures}

\end{document}